\documentclass[a4paper,11pt]{article}

\usepackage{pos}
\usepackage{bm}

\title{Towards low-dimensionalization of 
four-dimensional QCD}
\ShortTitle{Towards low-dimensionalization of 
four-dimensional QCD}

\author*{Kei Tohme}
\author{Hideo Suganuma}

\affiliation{Department of Physics, Graduate School of Science, Kyoto University, \\
Kitashirakawa-oiwake, Sakyo, Kyoto 606-8502, Japan}

\emailAdd{tohme@ruby.scphys.kyoto-u.ac.jp}
\emailAdd{suganuma@scphys.kyoto-u.ac.jp}

\abstract{
 Inspired by the one-dimensional color-electric flux-tube in a hadron, 
 we propose a possible way of low-dimensionalization of 4D QCD.
 As a strategy, we use gauge degrees of freedom and propose a new gauge fixing of ``dimensional reduction (DR) gauge".
 The DR gauge is defined so as to minimize
 $R_{\rm DR} \equiv \int d^4s~ 
 {\rm Tr}~[A^2_x(s)+A^2_y(s)]$,
 which preserves the 2D SU($N_{c}$) gauge symmetry. 
 We investigate low-dimensionalization properties of 4D DR-gauged QCD in
 SU(3) lattice QCD at $\beta$ = 6.0.
 In the DR gauge, the amplitudes of two gluon components $A_{x}(s)$ and $A_{y}(s)$ are found to be strongly suppressed,
 and these components have a large effective mass of $M_{\perp} \simeq 1.7$ GeV.
In the DR gauge, 
the static interquark potential is well
reproduced only with the two components $A_{t}(s)$ and $A_{z}(s)$, while $A_{x}(s)$ and $A_{y}(s)$ 
seem to be inactive. 
 We investigate the spatial correlation of two 
 $t$-directed gluons and find that the correlation decreases as $e^{-mr}$ with $m \simeq$ 0.64 GeV, 
 corresponding to the correlation length $\xi \equiv 1/m \simeq$ 0.31 fm.
 We reduce 4D QCD in the DR gauge to 2D QCD with  the coupling $g_{2D} = g/\xi$, 
 which approximately reproduces the string tension.
}
\FullConference{%
The XVIth Quark Confinement and 
the Hadron Spectrum Conference 
(QCHSC24), 
19-24 August, 2024, Cairns Convention Centre, Cairns, Queensland, Australia
}

\begin{document}
\maketitle

\section{Introduction}
\label{sec:intro}

Quantum chromodynamics (QCD) is the fundamental theory of the strong interactions,
and its analytical solving is important to understand the vacuum structure and non-perturbative phenomena in QCD.
However, it is a difficult problem due to its strong-coupling nature in the low-energy region.

Quark confinement is a typical example of non-perturbative QCD phenomena
and understanding its mechanism
is one of the most challenging problems in four-dimensional (4D) QCD.
Although it has not yet been solved in an analytical manner, 
it is known that this phenomenon is characterized by a linear interquark potential \cite{CREUTZ_MCSU2_1980}
and one-dimensional squeezing of color-electric fields \cite{Handbook_2023}.
Compared with the ordinary waves (e.g., electromagnetic fluxes)
which spread over three-dimensional space,
this one-dimensional color-flux squeezing is a singular phenomenon.
This can be regarded as a reduction of the spatial dimension by two and
can be considered as a kind of low-dimensionalization.

A famous example of low-dimensionalization is related to a random magnetic system.
Indeed, G.~Parisi and N.~Sourlas showed the equivalence of a $d$-dimensional spin system under Gaussian random magnetic fields and the $(d-2)$-dimensional system without magnetic fields (the Parisi-Sourlas mechanism) \cite{Parisi_Sourlas_1979}.
The QCD vacuum also has a random structure of magnetic fields.
In 1977, G.K.~Savvidy pointed out that the QCD vacuum is filled with color magnetic fields \cite{Savvidy_1977},
and J.~Ambjorn and P.~Olesen found that the magnetic fields have a random structure at a large scale, which is called the Copenhagen (spaghetti) vacuum \cite{Ambjorn_Olesen_1980}.
Thus, there might be some connection between the low-dimensionalization in 4D QCD and the Parisi-Sourlas mechanism, where random magnetic fields play an important role \cite{Iritani_Suganuma_Iida_2009}.

As another scenario of the low-dimensionalization of QCD, 
J.~Greensite and Š.~Olejník studied 
quark confinement via dimensional reduction 
in 2+1 dimensional Yang-Mills theory, 
assuming an approximate form for 
the ground-state vacuum in the temporal gauge
\cite{DR_Greensite}.

In any case, ``low-dimensionalization'' might be a key concept in 4D QCD 
and give some hints for understanding the mechanism of quark confinement.
Therefore, we study the low-dimensionalization properties of 4D QCD in this paper.
In particular, we explore the possibility of describing 4D QCD in terms of 2D QCD-like degrees of freedom.
As a strategy, we use gauge degrees of freedom and propose a new gauge fixing of ``dimensional reduction (DR) gauge'' \cite{TS24}.

This paper is organized as follows.
In Sec. \ref{sec:DRG}, we formulate the DR gauge and $tz$-projection in both continuous and lattice spacetime.
The lattice QCD calculations and numerical analyses in the DR gauge are performed in Sec. \ref{sec:numerical_calculation}.
In Sec. \ref{sec:2D_model}, we discuss on a 2D modeling of 4D QCD in the DR gauge with an approximation.
Section \ref{sec:summary} is devoted for the summary and concluding remarks.

\section{Dimensional Reduction Gauge}
\label{sec:DRG}

\subsection{Formalism of DR gauge in continuum QCD}
\label{subsec:DR_cont}
We define the dimensional reduction (DR) gauge \cite{TS24} so as to {\it minimize} the following function,
\begin{eqnarray}
    \label{eq:DRG}
    R_{\mathrm{DR}}
    \equiv
    \int d^{4} s \hspace{-5pt}
    \sum_{\perp = x, y} \hspace{-5pt}
    \mathrm{Tr}\left[ 
    A_{\perp}(s)^2
    \right] 
    =
    \int \hspace{-2pt}
    d^{4} s \;
    \mathrm{Tr}\left[ 
    A_{x}(s)^2 + A_{y}(s)^2
    \right]
\end{eqnarray}
with the gauge transformation.
In this paper, we use the subscript ``$\perp$'' to represent $x$- and $y$-directions, i.e., $\perp = x, y$.
From the definition
\eqref{eq:DRG}, the amplitudes of $A_{x}(s)$ and $A_{y}(s)$ are considered to be suppressed in the DR gauge.
Since $R_{\mathrm{DR}}$ does not contain a temporal-component $A_{t}(s)$, the DR gauge can be defined in both Minkowski and Euclidean spacetime. 
The local condition of the DR gauge is written as
\begin{equation}
    \label{eq:DRG_local}
    \partial_{\perp} A_{\perp} = \partial_{x}A_{x} + \partial_{y} A_{y} = 0 \; .
\end{equation}
This is a low-dimensional version of
the Landau gauge ($\partial_{\mu}A_{\mu} = 0$)
or
the Coulomb gauge ($\bm{\nabla} \cdot \bm{A} = 0$).

The 4D Yang-Mills (YM) action in the DR gauge is expressed by
\begin{equation}
    \label{eq:DRG_action}
    S_{\mathrm{DR}} = 
    \int d^{4} s 
    \left[ 
    -
    \frac{1}{2}
    \mathrm{Tr} \; G_{\mu\nu}G^{\mu\nu}
    +
    \frac{1}{2\alpha}
    \sum_{\perp = x, y}
    \mathrm{Tr}
    \left( \partial_{\perp}A_{\perp} \right)^{2}
    \right] ,
\end{equation}
where $G_{\mu\nu} \equiv \partial_{\mu}A_{\nu} - \partial_{\nu}A_{\mu} + ig[A_{\mu}, A_{\nu}]$ is the field strength tensor with gauge coupling $g$.
The first term in \eqref{eq:DRG_action} is a usual gauge-invariant 4D YM action, and the second a gauge-fixing term for the DR gauge.
Despite the gauge-fixing term, the action \eqref{eq:DRG_action} has a residual gauge symmetry for a transformation with $\Omega(t,z) \in \mathrm{SU}(N_{c})$,
\begin{eqnarray}
    \label{eq:res_sym_tz}
    A_{t,z}(s) 
    & \to &
    \Omega(t,z)
    \left( 
    A_{t,z}(s) + \frac{1}{ig}\partial_{t,z}
    \right)
    \Omega^{\dagger}(t,z) , \\
    \label{eq:res_sym_xy}
    A_{\perp}(s) 
    & \to &
    \Omega(t,z)
    \left( 
    A_{\perp}(s) + \frac{1}{ig}\partial_{\perp}
    \right)
    \Omega^{\dagger}(t,z) 
    = 
    \Omega(t,z)
    A_{\perp}(s)
    \Omega^{\dagger}(t,z) .
\end{eqnarray}
The action \eqref{eq:DRG_action} is invariant under the gauge transformation,
and this residual symmetry is the same as the gauge symmetry of 2D QCD on a $t$-$z$ plane.

To explore the possibility of a 2D description of 4D QCD in terms of $A_{t}(s)$ and $A_{z}(s)$, 
we introduce ``$tz$-projection'' 
as the removal of 
$x$- and $y$-directed gluons, $A_{x}(s)$ and $A_{y}(s)$, 
from the gauge configuration generated in lattice QCD \cite{TS24}.
In fact, the $tz$-projection is defined as follows,
\begin{eqnarray}
    \label{eq:tz_proj}
    \hspace{-90pt}
    tz\textrm{-projection :}
    \hspace{25pt}
    A_{x,y}(s) \to 0 .
\end{eqnarray}
Applying the $tz$-projection, 
the 4D DR-gauged QCD action  \eqref{eq:DRG_action} becomes
\begin{eqnarray}
    \label{eq:DRG_act_tzp}
    S^{tz}_{\mathrm{DR}} 
    = 
    \int dx dy
    \int dt dz \;
    \left[ 
    \mathrm{Tr} \; G^{2}_{tz}  
    + \!
    \sum_{\perp=x,y}
    \mathrm{Tr} 
    \left\{
      \left( \partial_{\perp}A_{t} \right)^2
    - \left( \partial_{\perp}A_{z} \right)^2
    \right\}
    \right]
\end{eqnarray}
at the tree level.
Ignoring the integral with respect to $x$ and $y$,
the first term in 
\eqref{eq:DRG_act_tzp} is the same as the 2D YM action on a $t$-$z$ plane.
The second term represents a deviation from the 2D YM action,
which is interpreted as $x$- or $y$-directed derivative interaction between 
2D systems.

Thus, after the $tz$-projection \eqref{eq:tz_proj}, 4D DR-gauged QCD can be expressed as 2D QCD-like systems on $t$-$z$ planes.
These $t$-$z$ planes are cross sections of 4D continuous spacetime, and there is the $x$- or $y$-directed derivative interaction between these cross sections.

\subsection{Lattice QCD formalism for DR gauge}
\label{subsec:DR_lat}

We formulate $\mathrm{SU}(N_{\rm{c}})$ lattice QCD on a 4D lattice with spacing $a$ in the Euclidean spacetime.
In lattice QCD, the gluon field is described as the link-variable $U_{\mu}(s) \equiv e^{iag A_{\mu}(s)} \in \mathrm{SU}(N_{\rm{c}})$, instead of $A_{\mu}(s) \in \mathfrak{su}(N_{\rm{c}})$.
We use the standard plaquette action
$S^{\mathrm{lat}} 
    \equiv 
    \beta \sum_{s} \left[
    1 - \frac{1}{N_{c}}\sum_{\mu < \nu} \mathrm{ReTr} \; P_{\mu\nu}(s)
    \right]$
with $\beta \equiv 2N_{c}/g^{2}$ and the plaquette variable,
$P_{\mu\nu}(s)
    \equiv
    U_{\mu}(s)
    U_{\nu}(s+a_{\mu})
    U^{\dagger}_{\mu}(s+a_{\nu})
    U^{\dagger}_{\mu}(s).$
Here, $a_{\mu}$ is the $\mu$-directed four vector of length $a$.

\begin{figure}[h]
\begin{center}
    \includegraphics[width=7.0cm]{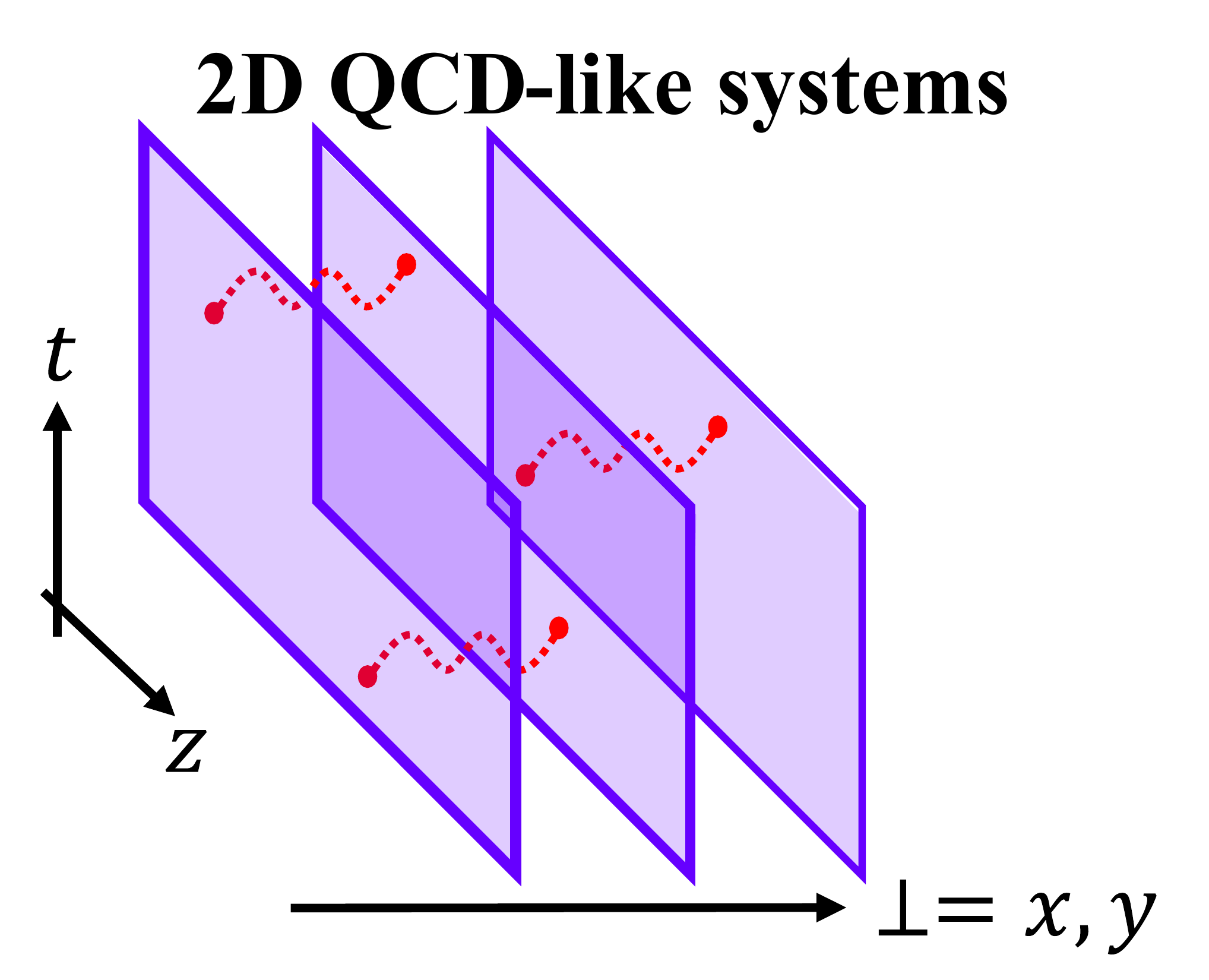}
\end{center}
\vspace{-15pt}
\caption{
    Schematic figure of 
    the reduction of 4D QCD into 
    2D QCD-like systems 
    under the $tz$-projection in the DR gauge. 
    The $tz$-projected 4D QCD system 
    is reduced to 2D QCD-like systems piled in the $x$- and $y$-directions.
    These 2D QCD-like systems interact with their neighbors, as shown by the wavy dotted lines.
    }
    \label{fig:tz_proj.ed_QCD4_4d}
\end{figure}

In lattice QCD, the DR gauge is defined so as to {\it maximize} 
\begin{eqnarray}
    \label{eq:DRG_lat}
    R^{\mathrm{lat}}_{\mathrm{DR}}
    \equiv
    \sum_{s}
    \sum_{\perp = x, y}
    \mathrm{Re Tr}
    \left[
        U_{\perp}(s)
    \right]
    = 
    \sum_{s}
    \mathrm{Re Tr}
    \left[
        U_{x}(s) + U_{y}(s)
    \right]
\end{eqnarray}
with gauge transformation.
The variation of $R^{\mathrm{lat}}_{\mathrm{DR}}$ for local gauge transformation is expressed with
\begin{equation}
    \label{eq:variation_DR_lat}
    \Delta_{\mathrm{DR}}(s)
    \equiv 
    \sum_{\perp = x, y}
    \partial^{\mathrm{B}}_{\perp} \bar{A}_{\perp}(s) ,
    \hspace{10pt}
    \bar{A}_{\perp}(s)
    \equiv 
    \frac{1}{2iag} 
    \left.
    \left[
        U_{\perp}(s) - U^{\dagger}_{\perp}(s) 
    \right]
    \right|_{\mathrm{traceless}} 
    \in
    \mathfrak{su}(N_{c}) ,
\end{equation}
where $\partial^{\mathrm{B}}_{\perp}$ is the backward derivative in the $\perp$-direction.
$\bar{A}_{\perp}(s)$ is a lattice gluon field,
and the subscript ``traceless'' means the subtraction of its trace.  
Taking the continuum limit ($a \to 0$), this lattice definition with \eqref{eq:DRG_lat} leads to the continuum definition of the DR gauge \eqref{eq:DRG_local}
since the backward derivative becomes the usual derivative.
The $tz$-projection on a lattice is defined as
\begin{equation}
    \label{eq:tz_proj_lat}
    \hspace{-90pt}
    tz\textrm{-projection :}
    \hspace{28pt}
    U_{x,y}(s) \to 1 . 
\end{equation}
Then, the $tz$-projected plaquette action becomes
\begin{eqnarray}
    \label{eq:DRG_lat_act_tzp}
    && \hspace{-20pt}
    S^{\mathrm{lat}}_{tz\mathrm{-DR}} 
    = 
    \beta \sum_{s}
    \left[
    \left\{
    1 - \frac{1}{N_{c}} \mathrm{ReTr}P_{tz}(s)
    \right\} 
    + 
    \sum_{\mu = t, z} \hspace{-3pt}
    \left\{
        1 - \frac{1}{N_{c}} \hspace{-2pt}
        \sum_{\perp = x,y} \hspace{-5pt}
        \mathrm{ReTr} \hspace{-2pt}
        \left[
            U_{\mu}(s) U^{\dagger}_{\mu}(s+a_{\perp})
        \right]
    \right\}
    \right] .
\end{eqnarray}
Similarly in the continuum case, the first term of 
\eqref{eq:DRG_lat_act_tzp} is the plaquette lattice action of the 2D YM theory, and the second term the interaction between neighboring $t$-$z$ planes in terms of link-variables.
Thus, in lattice QCD, the $tz$-projection \eqref{eq:tz_proj_lat} reduces the 4D DR-gauged YM theory to 2D QCD-like systems on $t$-$z$ planes as shown in Fig.\ref{fig:tz_proj.ed_QCD4_4d}.
These 2D systems are piled in the $x$- and 
$y$-directions, and they interact with their neighboring planes in these directions.

\section{Numerical Calculation}
\label{sec:numerical_calculation}
To investigate non-perturbative properties of the 4D DR-gauged YM theory, we perform $\mathrm{SU}(N_{\rm{c}})$ lattice QCD simulations at the quenched level.
We use the standard plaquette action with $\beta = 6.0$ and $L^{4} = 24^{4}$, 
corresponding to the lattice spacing $a \simeq 0.1 \; \mathrm{fm}$
and the total volume $L^{4} \simeq (2.4 \; \mathrm{fm})^{4}$.
We use the pseudo-heat bath algorithm and generate 800 gauge configurations, which are picked up every 1,000 sweeps after 20,000 sweeps for thermalization.

After the importance sampling, we perform the DR gauge fixing by numerically maximizing the function \eqref{eq:DRG_lat} for each configuration using the iterative maximization algorithm, similar to 
the setting of the Landau/Coulomb gauge fixing \cite{Iritani_Suganuma_Iida_2009, Iritani_Suganuma_generalized_landau}.
To achieve fast convergence of the algorithm, we adopt the over-relaxation (OR) method with the OR parameter 1.6.
When $R^{\mathrm{lat}}_{\mathrm{DR}}$ is maximized, 
the variation \eqref{eq:variation_DR_lat} has to be zero.
As a numerical convergence criterion, we impose that the maximization of $R^{\mathrm{lat}}_{\mathrm{DR}}$ stops when 
$
    \epsilon_{\mathrm{DR}}
    \equiv 
    \frac{1}{N_{c} N_{s}}
    \sum_{s}
    \mathrm{Tr}
    \left[
        \Delta_{\mathrm{DR}}(s)
        \Delta^{\dagger}_{\mathrm{DR}}(s)   
    \right]
    < 4.0 \times 10^{-12}
$
is satisfied.

The vacuum expectation value (VEV) is estimated using the Monte Carlo method and the error is estimated with the jackknife method.
We use $\langle \cdots \rangle$ as a usual VEV and $\langle \cdots \rangle_{\mathrm{DR}}$ as a VEV calculated with DR-gauged configurations.

\subsection{Suppression of gauge amplitudes in the DR gauge}
To begin with, we investigate the suppression of gauge amplitudes $A_{x}(s)$ and $A_{y}(s)$ in the DR gauge.
We calculate a distance between a link-variable $U_{\perp}$ ($\perp = x,y$) and the unit matrix $I$
since $A_{\perp} = 0$ corresponds $U_{\perp} \equiv e^{iag A_{\perp}} = I$.
Denoting the distance between matrices $A$ and $B$ as $d(A, B)$, we define the squared distance as 
$
    d(A, B)^{2} 
    \equiv
    \frac{1}{2N_{c}} \mathrm{Tr}\left[ 
    (A-B)^{\dagger}
    (A-B)
    \right]
$
, which is proportional to the Frobenius norm of the matrix $(A-B)$ \cite{TS24}.

The distance between an $x$- or $y$-directed link variable $U_{\perp}$ and a unit matrix $I$ is calculated as
\begin{equation}
    \label{eq:distance_DR}
    \langle d(U_{\perp}, I)^{2} \rangle_{\mathrm{DR}}
    =
    1 - \frac{1}{N_{c}} \mathrm{ReTr} \; U_{\perp}
    \simeq
    0.076 .
\end{equation}
Since
$
\langle d(U_{\mu}, I)^{2} \rangle
=
1 - \frac{1}{N_{c}} \mathrm{ReTr} \, U_{\mu}
\in \left[0, \frac{3}{2} \right]
$
holds in SU(3) case \cite{TS24},
the result \eqref{eq:distance_DR} shows a strong suppression of the amplitudes of  $A_{x}(s)$ and $A_{y}(s)$ in the DR gauge.

\subsection{Spatial correlation of spatial-links in DR gauge}
\label{subsec:cor_spatial}
In the previous subsection, it is found that the amplitudes of $A_{\perp}(s) \; (\perp = x,y)$ are strongly suppressed in the DR gauge.
Next, we consider the spatial correlation of $A_{\perp}(s)$ in the $\perp$-direction.

We define the spatial correlation of two $\perp$-directed link-variables 
$U_{\perp}(s)$
in the DR gauge as 
\begin{equation}
    \label{eq:UxVx_x}
    \hspace{-2pt}
    F(r)
    \equiv
    \frac{1}{N_{c}}
    \langle
    \mathrm{Tr} \; U_{\perp}(0) U^{\dagger}_{\perp}(ra_{\perp})
    \rangle_{\mathrm{DR}}  
    =
    \frac{a^{2}}{\beta}
    \langle
    A^{a}_{\perp}(0) A^{a}_{\perp}(r a_{\perp}) 
    \rangle_{\mathrm{DR}}
    +
    \left\{
        1 - \frac{a^{2}}{\beta} \langle A^{a}_{\perp}(0)^{2} \rangle_{\mathrm{DR}}
    \right\}
    +
    O(a^{4}) ,
\end{equation}
where $\perp$ takes $x$ or $y$.
The right hand side is derived from the expansion of link-variables with respect to gauge fields,
and its first term 
is the gluon propagator.
We calculate this correlation $F(r)$ 
in SU(3) lattice QCD.

\begin{figure}[h]
    \centering
    \includegraphics[width=8.0cm]{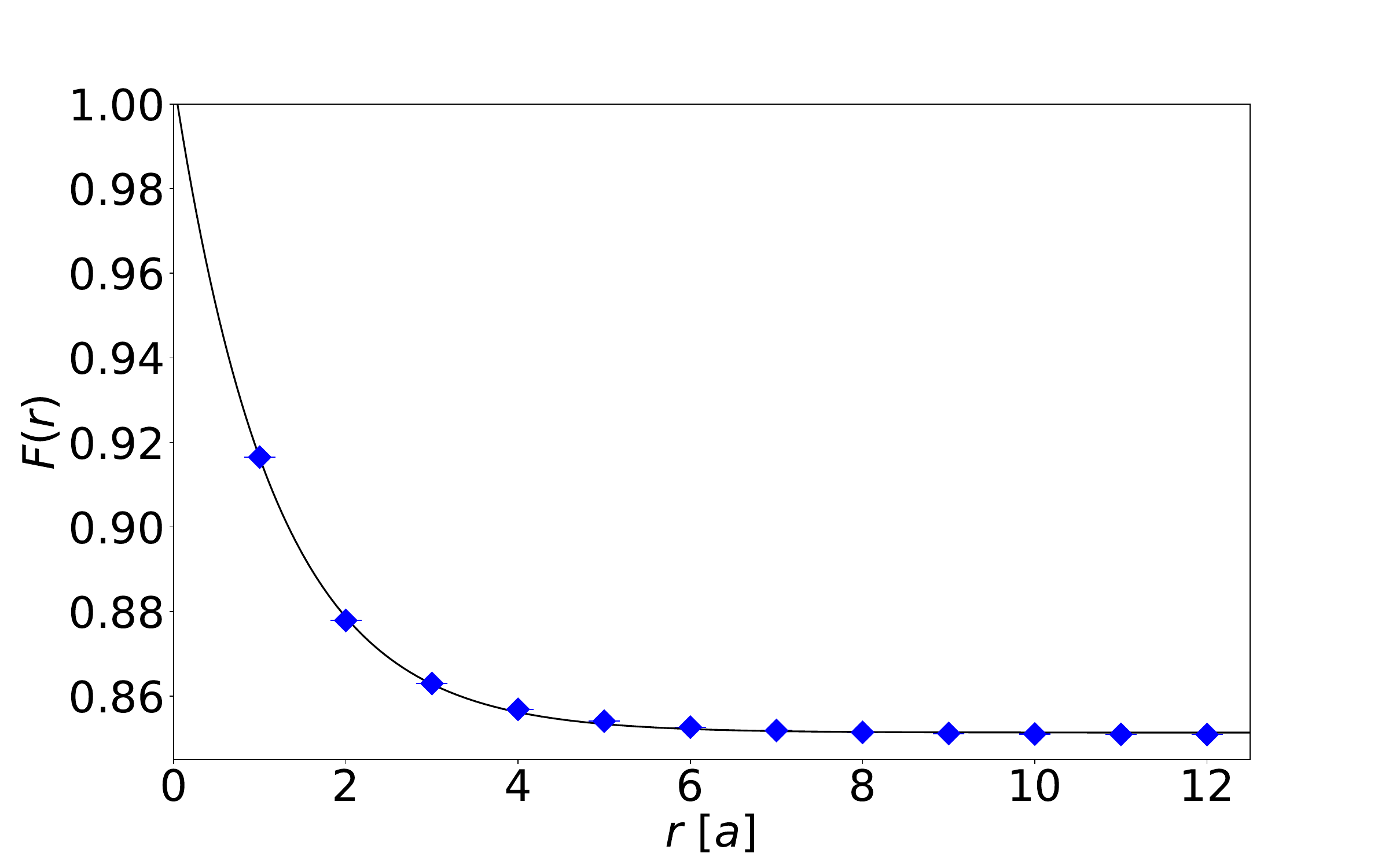}
    \vspace{-6pt}
    \caption{
    The spatial correlation of $\perp$-directed link-variables, $F(r) \equiv \frac{1}{N_{c}} \langle \mathrm{Tr} ~ U_{\perp}(0)U^{\dagger}_{\perp}(r a_{\perp}) \rangle_{\mathrm{DR}}$ ($\perp = x,y$), in lattice QCD at $\beta = 6.0$.
    The dots are the lattice data, and the solid line the best exponential fit $A e^{-M_{\perp}r} + B$ with $A \simeq 0.155$, $M_{\perp} \simeq 1.71 \; \mathrm{GeV}$ and $B \simeq 0.851$.
    This figure is taken from Ref. \cite{TS24}.
    }
    \label{fig:UxVx_x}
\end{figure}

Figure \ref{fig:UxVx_x} shows the lattice QCD result for $F(r)$ at $\beta = 6.0$. The lattice data denoted by dots are well reproduced with the exponential function
\begin{equation}
    \label{eq:product_UxVx_x}
    F(r)
    \simeq
    A e^{-M_{\perp}r} + B \,, \; \; \;
    A \simeq 0.155, \;\;
    M_{\perp} \simeq 0.87a^{-1}, \;\;
    B \simeq 0.851 .
\end{equation}
The behavior of the gluon propagator is described with $A$ and $M_{\perp}$, and a constant of the second term in Eq. \eqref{eq:UxVx_x} corresponds to $B$.

The propagation of $A_{\perp}(s)$ in the $\perp$-direction is exponential.
The behavior is controlled by the parameter $M_{\perp}$, which can be regarded as the effective mass of $A_{\perp}(s)$.
Thus, the spatial effective mass of $A_{\perp}(s)$ is estimated to be $M_{\perp} \simeq 1.71 \; \mathrm{GeV}$, 
and $x$- and $y$-components of gluon fields seem 
to be massive in the DR gauge.
Thus, $A_{x}(s)$ and $A_{y}(s)$ are inactive in the infrared region, and therefore $A_{t}(s)$ and $A_{z}(s)$ are considered to be dominant for quark confinement in the DR gauge.

\vspace{-13pt}

\vspace{-1pt}

\subsection{Wilson loops in DR-gauged QCD}
\label{sec:Wilson_loop}
In the DR gauge, $A_{x}(s)$ and $A_{y}(s)$ seem to have a large effective mass, and therefore we expect them to be inactive in the infrared region and the other components, $A_{t}(s)$ and $A_{z}(s)$, to be dominant for quark confinement.
This conjecture is inspired by confinement studies 
in the maximally abelian (MA) gauge, 
which keeps partial $\mathrm{U}(1)^{N_{c}-1}$ gauge symmetry and 
strongly suppresses the amplitudes of off-diagonal gluon components $A^{\alpha}_{\mu}(s)$ \cite{Ichie_Suganuma_1999, Sakumichi_Suganuma_AD_2015}. 
In the MA gauge, 
$A^{\alpha}_{\mu}(s)$ have a large effective mass of $M_{\mathrm{off}} \simeq 1 \; \mathrm{GeV}$ \cite{Amemiya_Suganuma_1999, Gongyo_Iritani_Suganuma_2012}, 
and the string tension 
is perfectly reproduced only with the diagonal components \cite{Sakumichi_Suganuma_AD_2015}. 
From the analogies summarized in Table~\ref{tab:comparison}, the low-energy phenomena in 4D QCD might be described only with $A_{t}(s)$ and $A_{z}(s)$ in the DR gauge.
To examine this conjecture, 
we investigate Wilson loops 
with applying the $tz$-projection  
in the DR gauge in lattice QCD \cite{TS24}. 

\begin{center}
\vspace{-3.8pt}
\begin{table}[h]
    \centering
    \begin{tabular}{c|c|c} 
    \hline
      & DR gauge & MA gauge  \\ 
    \hline
    Minimized functional 
    &
    $R_{\mathrm{DR}}
    \equiv
    \int d^{4} s 
    \sum_{\perp = x, y}
    \mathrm{Tr}\left[ 
    A_{\perp}(s)^2
    \right]$
    &
    $R_{\mathrm{MA}}
    \equiv
    \int d^{4} s
    \sum_{\mu,\alpha}
    \left[ 
    A^{\alpha}_{\mu}(s) A^{-\alpha}_{\mu}(s)
    \right]$
    \\
    Residual symmetry
    &
    2D $\mathrm{SU}(N_{c})$ gauge symmetry
    &
    4D $\mathrm{U}(1)^{N_{c}-1}$ gauge symmetry
    \\
    Suppressed gluons
    &
    $A_{\perp} $ ($x$- and $y$-components)
    &
    $A^{\alpha}_{\mu}$ (off-diagonal components) \\
    Effective gluon mass 
    &
    $M_{\perp} \simeq 1.7 \; \mathrm{GeV}$
    &
    $M_{\mathrm{off}} \simeq 1 \; \mathrm{GeV}$ \\
    \hline
    \end{tabular}
    \caption{
    Comparison of the DR gauge and the MA gauge.
    Both gauges are defined in a similar manner to {\it minimize} some functional, 
    listed in the first row.
    In the DR gauge, 2D $\mathrm{SU}(N_{c})$ gauge symmetry remains, 
    the amplitudes of $x$- and $y$-components $A_\perp(s)$ of gluons are suppressed, 
    and $A_\perp(s)$ have a large effective mass of $M_{\perp} \simeq 1.7 \; \mathrm{GeV}$.
    In the MA gauge, 4D $\mathrm{U}(1)^{N_{c}-1}$ gauge symmetry remains, 
    the amplitudes of off-diagonal gluons 
    $A_{\mu}^{\alpha}(s)$ are suppressed, and $A_{\mu}^{\alpha}(s)$ have a large effective mass of $M_{\mathrm{off}} \simeq 1 \; \mathrm{GeV}$ \cite{Amemiya_Suganuma_1999, Gongyo_Iritani_Suganuma_2012}.
    }
    \label{tab:comparison}
\end{table}
\vspace{-10pt}
\end{center}

\subsubsection{$tz$-projected Wilson loop}

First, we calculate the $tz$-projected Wilson loop $\langle W^{tz}(r,T)\rangle_{\mathrm{DR}}$ \cite{TS24} after 
the DR gauge fixing and applying 
the $tz$-projection of  $A_{x,y}\rightarrow 0$ for  
the gauge configuration obtained in lattice QCD.
Since all loops on the $t$-$z$ plane are trivially unaffected by the $tz$-projection \eqref{eq:tz_proj_lat}, 
we focus on Wilson loops on $t$-$\perp \, (\perp=x,y)$ planes. 
The $tz$-projected interquark potential $V^{tz}(r)$ is 
obtained as
\vspace{-1pt}
\begin{equation}
    \label{eq:potential_tz_proj}
    \langle W^{tz}(r,T) \rangle_{\mathrm{DR}}
    =
    A e^{-V^{tz}(r)T}
\vspace{-2.5pt}
\end{equation}
in the accurate lattice calculation 
using the gauge-covariant smearing method \cite{APE_smearing_1987, Takahashi_Suganuma_detailedQQpot_2002}.

\begin{figure}[h]
    \begin{minipage}[h]{0.48\columnwidth}
        \centering
        \includegraphics[width=7.2cm]{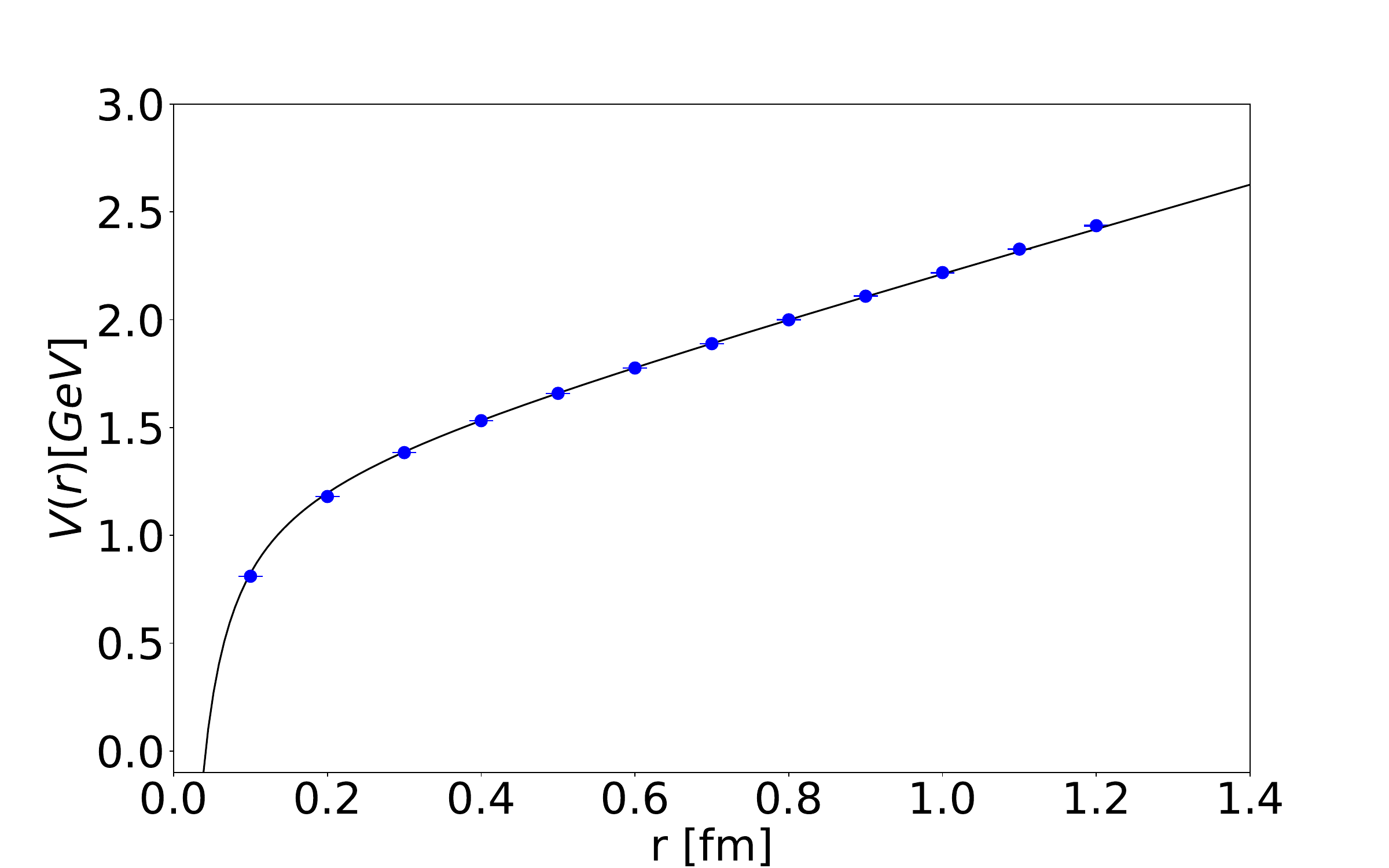}
        \caption{
        The $tz$-projected interquark potential $V^{tz}(r)$, calculated from 
        $\langle W^{tz}(r,T) \rangle_{\mathrm{DR}}$
        on the $t$-$\perp$ plane in lattice QCD. The solid line denotes the original static intrequark potential. 
        Taken from Ref.~\cite{TS24}.
        }
        \label{fig:wilson_tzp}
    \end{minipage}
    \hspace{20pt}
    \begin{minipage}[h]{0.48\columnwidth}
        \centering
        \includegraphics[width=7.3cm]{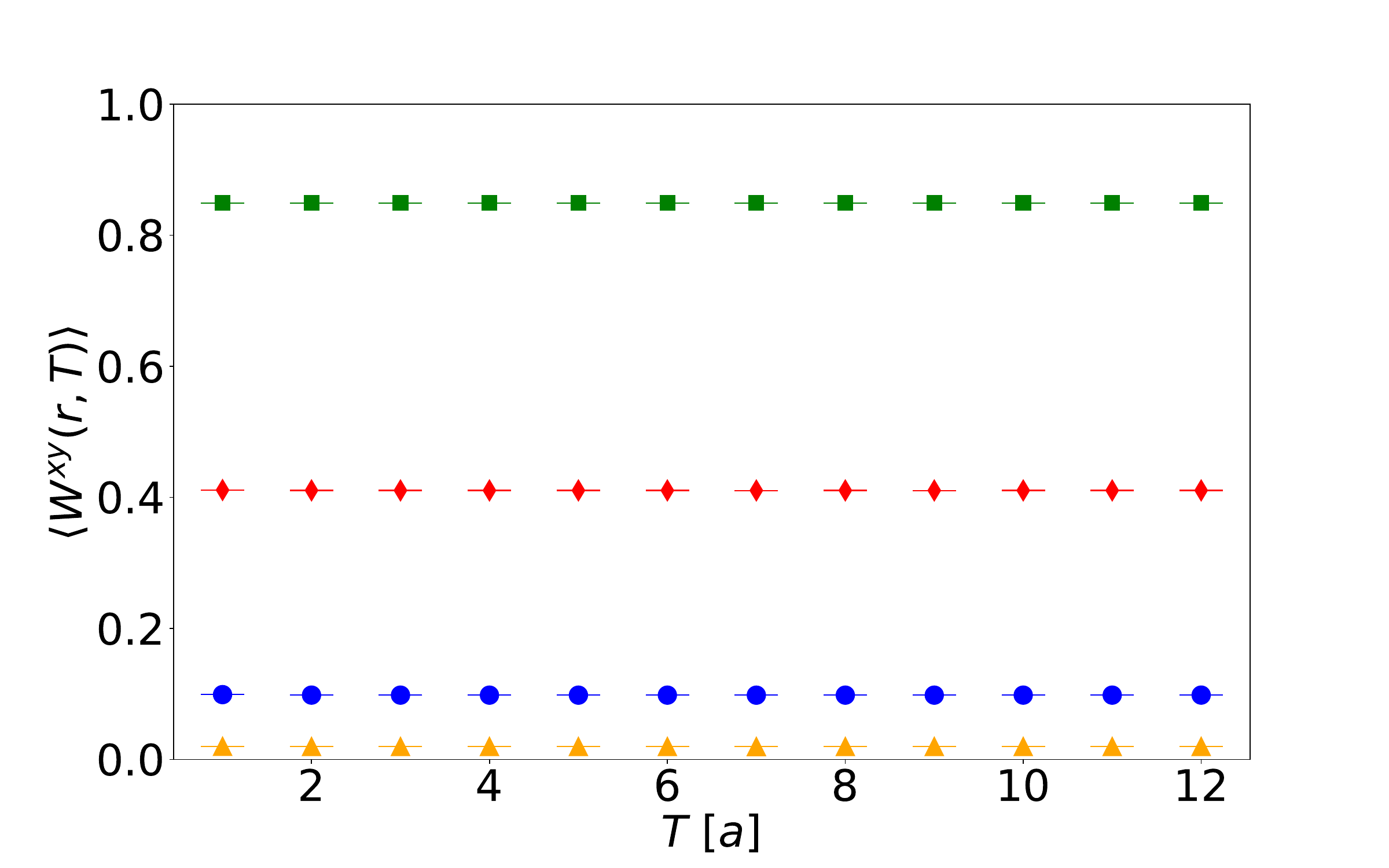}
        \vspace{-20pt}
        \caption{
        $T$-dependence of the $xy$-projected Wilson loop 
        $\langle W^{xy}(r, T) \rangle_{\mathrm{DR}}$ 
        in the DR gauge 
        for several $r$, i.e.,  
        $r=1$ (green square), $r=3$ (red diamond), $r=6$ (orange circle) and $r=9$ (orange triangle)
        \cite{TS24}.
        }
        \label{fig:wilson_xyp}
    \end{minipage}
\end{figure}

Figure~\ref{fig:wilson_tzp} shows the lattice QCD result of the $tz$-projected interquark potential $V^{tz}(r)$ in the DR gauge, plotted against 
the interquark distance $r$.
The dots denote the $tz$-projected potential $V^{tz}(r)$ in the DR gauge, and the solid line the original interquark potential calculated in $\mathrm{SU}(3)$ lattice QCD \cite{Takahashi_Suganuma_detailedQQpot_2002}.
The $tz$-projected potential $V^{tz}(r)$ is in good agreement with the original potential. 
This result shows that the interquark potential is reproduced 
only with  $A_{t}(s)$ and $A_{z}(s)$ 
in the DR gauge.

\vspace{-10pt}

\vspace{-6pt}
\subsubsection{$xy$-projected Wilson loop}

Next, we examine the contribution of $A_{\perp}(s)$ to quark confinement.
As an opposite operation to the $tz$-projection, we define ``$xy$-projection'' by the replacement 
of $A_{t}(s)$ and $A_{z}(s)$ by zero,
\begin{equation}
    \label{eq:xy_proj}
    \hspace{-90pt}
    xy\textrm{-projection :}
    \hspace{25pt}
    A_{t,z} \to 0 \:
    \Leftrightarrow  \:
    U_{t,z} \to 1,
\end{equation}
for the gauge configuration obtained in lattice QCD.
We apply the $xy$-projection to the gauge configurations and calculate the $xy$-projected Wilson loop  $\langle W^{xy}(r,T) \rangle_{\mathrm{DR}}$.

Figure~\ref{fig:wilson_xyp} shows
the $xy$-projected Wilson loop 
$\langle W^{xy}(r, T) \rangle_{\mathrm{DR}}$ 
plotted against the temporal length~$T$, 
and $\langle W^{xy}(r,T) \rangle_{\mathrm{DR}}$ 
is found to be independent of $T$.
Similarly to the ordinary potential, 
we define the $xy$-projected interquark potential $V^{xy}(r)$ from 
$\langle W^{xy}(r, T) \rangle_{\rm DR}$ 
and find it to be zero, 
\begin{equation}
\label{eq:potential_xy_proj}
    V^{xy}(r) 
    \equiv
    - \lim_{T \to \infty}
    \frac{1}{T}
    \ln 
    \langle
        W^{xy}(r, T)
    \rangle_{\mathrm{DR}}=0,
\end{equation}
because $\langle W^{xy}(r, T) \rangle_{\rm DR}$ is a $T$-independent constant for each $r$, as shown in Fig.~\ref{fig:wilson_xyp}.

This result seems to indicate a tiny contribution of $A_{x}(s)$ and $A_{y}(s)$ to quark confinement in the DR gauge.
Although they might make some contribution in the full (non-projected) Wilson loop due to the non-Abelian property of link-variables,
their contribution is considered to be negligible.

\subsection{Spatial correlation of temporal-links in DR gauge}
\label{subsec:correlation_temporal}
The results in the previous subsection show a dominant role of $A_{t}(s)$ and $A_{z}(s)$ for quark confinement in the DR gauge.
As an interesting possibility, low-energy phenomena in 4D DR-gauged QCD could be described only with 2D degrees of freedom, i.e., $A_{t}(s)$ and $A_{z}(s)$.
To examine this possibility, we here focus on the deviation between 2D QCD and $tz$-projected 4D QCD in the DR gauge.

\begin{figure}[h]
    \begin{minipage}[h]{0.45\columnwidth}
        \centering
        \includegraphics[width=6.0cm]{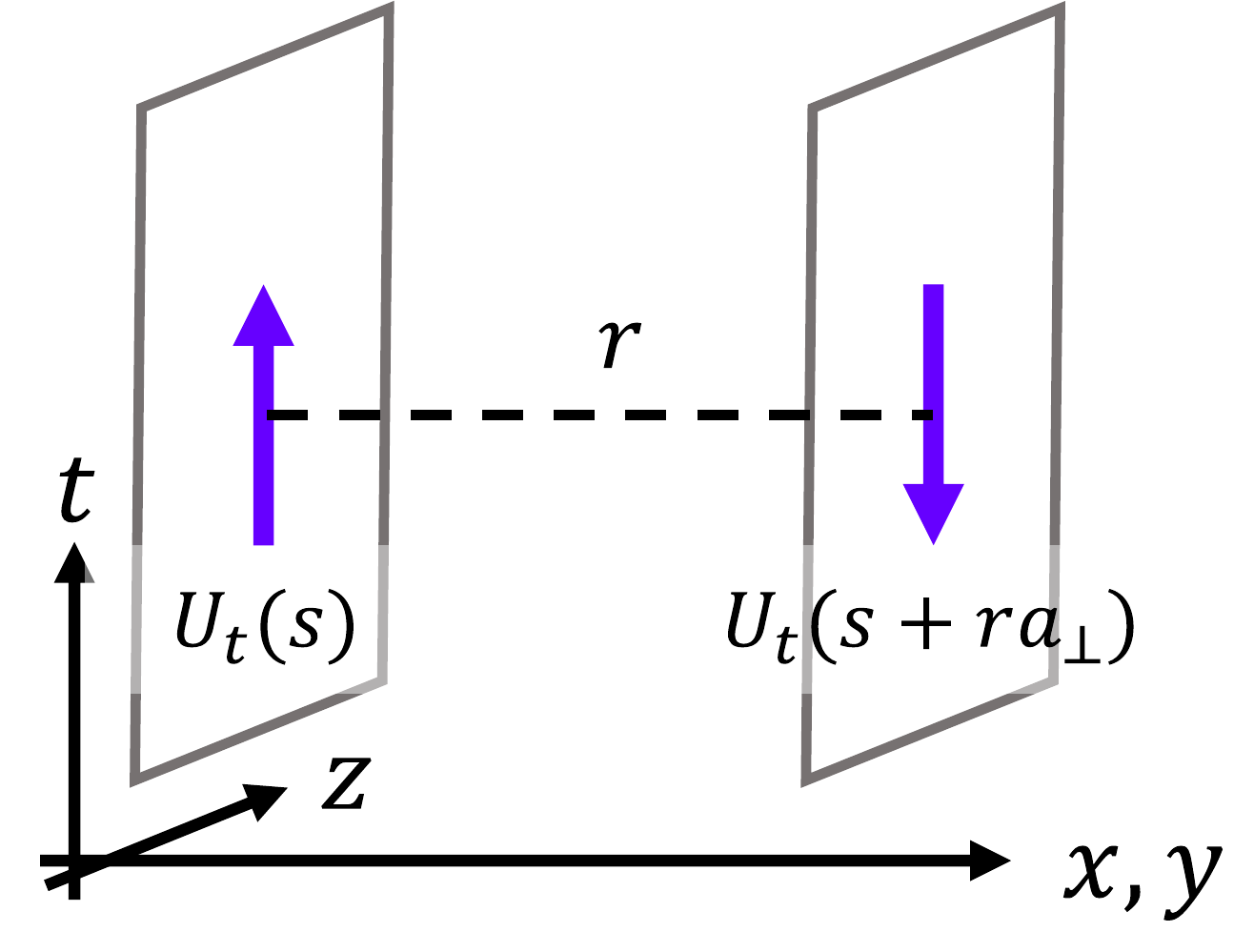}
        \caption{
        Schematic figure of the spatial correlation between link-variables $U_{t}(s)$ and $U^{\dagger}_{t}(s+r a_{\perp})$ on the $t$-$z$ planes separated by $r$ in the $\perp$-direction.
        }
        \label{fig:link_correlation}
    \end{minipage}
    \hspace{25pt}
    \begin{minipage}[h]{0.45\columnwidth}
        \includegraphics[width=7.0cm]{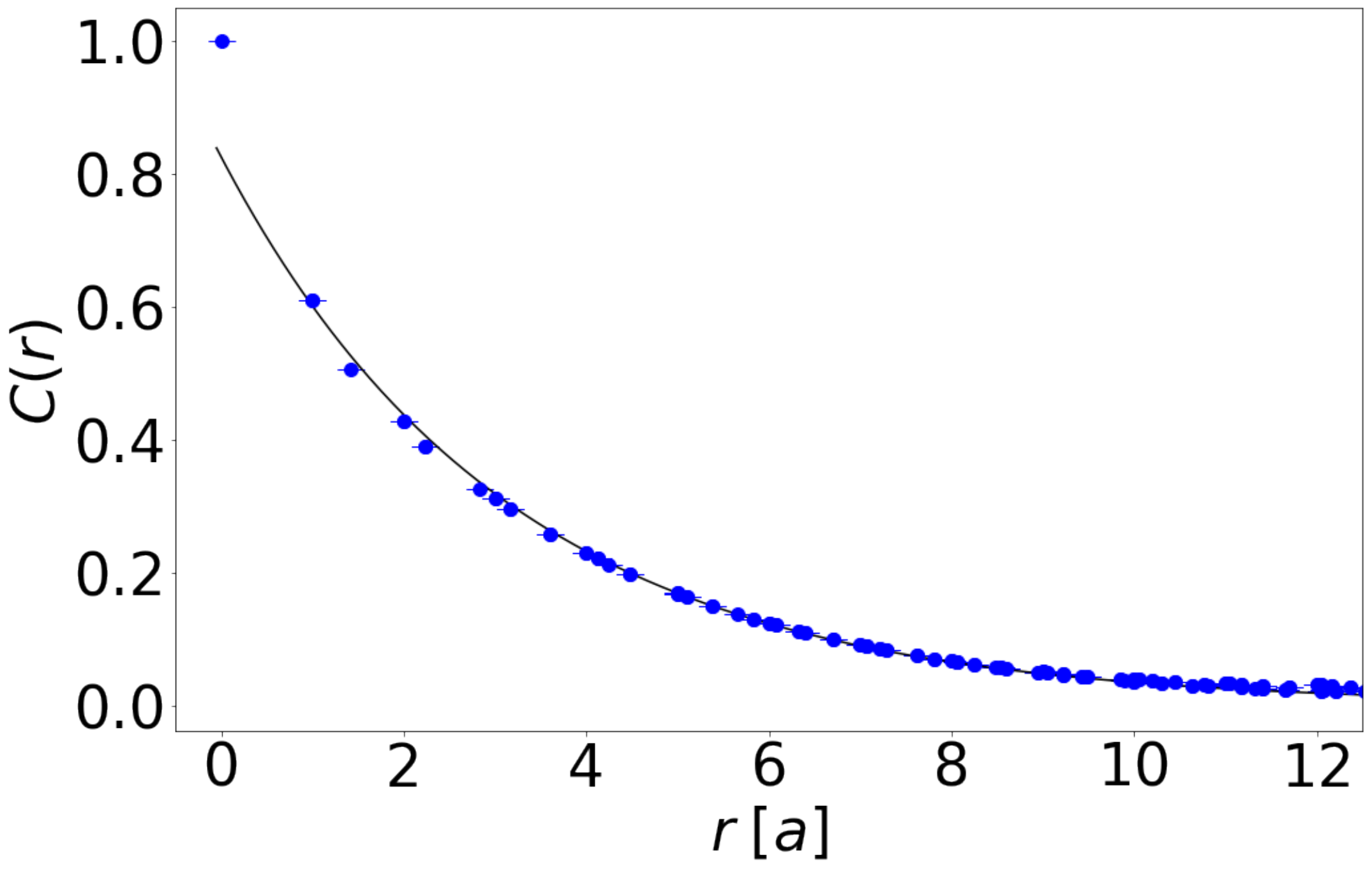}
        \caption{
        The $\perp$-directed spatial correlation 
        between two temporal link-variables
        in the DR gauge. 
        The dots are the lattice QCD data at $\beta = 6.0$, 
        and the solid line the best exponential fit $Ae^{-mr}$ 
        with $A \simeq 0.83$ and $m \simeq 0.64 ~ \mathrm{GeV}$.
        }
        \label{fig:U4cor}
    \end{minipage}
\end{figure}

The deviation between 2D QCD and $tz$-projected 4D QCD in the DR gauge is the additional derivative interaction in 
\eqref{eq:DRG_act_tzp}, which is written 
in lattice formalism 
as the second term of \eqref{eq:DRG_lat_act_tzp},
\begin{equation}
    \label{eq:QCD2_int}
    \beta \sum_{s} 
    \sum_{\mu = t, z}
    \left\{
        1 - \frac{1}{N_{c}}
        \sum_{\perp = x,y}
        \mathrm{ReTr} 
        \left[
            U_{\mu}(s) U^{\dagger}_{\mu}(s+a_{\perp})
        \right]
    \right\} .
\end{equation}
The $\perp$-directed local interaction induces a non-local correlation of $t$- or $z$-directed link-variables in the $\perp$-directions,
as schematically depicted in Fig. \ref{fig:link_correlation}.
This non-local correlation is unchanged by the exchange of $t$ and $z$ in Euclidean spacetime
since $t$ and $z$ are symmetric in this interaction \eqref{eq:QCD2_int}.

Then, using lattice QCD, 
we investigate 
the $\perp$-directed 
spatial correlation $C(r)$ 
of two 
temporal link-variables in the 
DR gauge: 
\begin{equation}
    \label{eq:link_correlation}
    C(r)
    \equiv
    \frac{1}{N_{c}}
    \langle 
    \mathrm{ReTr} \;
        U_{t}(s) 
        U^{\dagger}_{t}(s+r a_{\perp})
    \rangle_{\mathrm{DR}}
    =
    \frac{1}{N_{c}}
    \langle 
    \mathrm{ReTr} \;
        U_{z}(s) 
        U^{\dagger}_{z}(s+r a_{\perp})
    \rangle_{\mathrm{DR}}. 
\end{equation}
In this subsection, $a_{\perp}$ is used as a lattice unit vector on the $x$-$y$ plane, 
and both on-axis and off-axis correlations are calculated.

\vspace{-3pt}

Figure \ref{fig:U4cor} shows the $\perp$-directed spatial correlation $C(r)$ of two temporal link-variables
plotted against distance $r$ 
in the DR gauge in SU(3) lattice QCD at $\beta = 6.0$. 
This result is well reproduced 
by the exponential function 
\begin{equation}
    \label{eq:fit_curve_exp}
    C(r)
    \simeq
    A e^{-mr} ,\;\;\;
    A \simeq 0.83 , \;\;
    m \simeq 0.32 a^{-1} \simeq 0.64 \: \mathrm{GeV} .
\end{equation}
Thus, in the DR gauge, the propagation of $U_{t}(s)$ and $U_{z}(s)$ in the $\perp$-direction is massive with the effective mass of $m \simeq0.64~{\rm GeV}$, and their   correlation length $\xi$ in the $\perp$-direction is given by
\begin{equation}
    \label{eq:cor_length}
    \xi 
    \equiv
    \frac{1}{m}
    \simeq
    0.31 \; \mathrm{fm}.
\end{equation}
The spatial correlation $C(r)$ becomes small enough in larger region than the correlation length $\xi$.
Thus, the range of this $\perp$-directed correlation is approximately $\xi$, and it is a short-range correlation.

\section{Two-dimensional Modeling of Non-perturbative 4D QCD}
\label{sec:2D_model}
\begin{figure}[h]
    \begin{minipage}[h]{0.45\columnwidth}
        \centering
        \includegraphics[width=6.5cm]{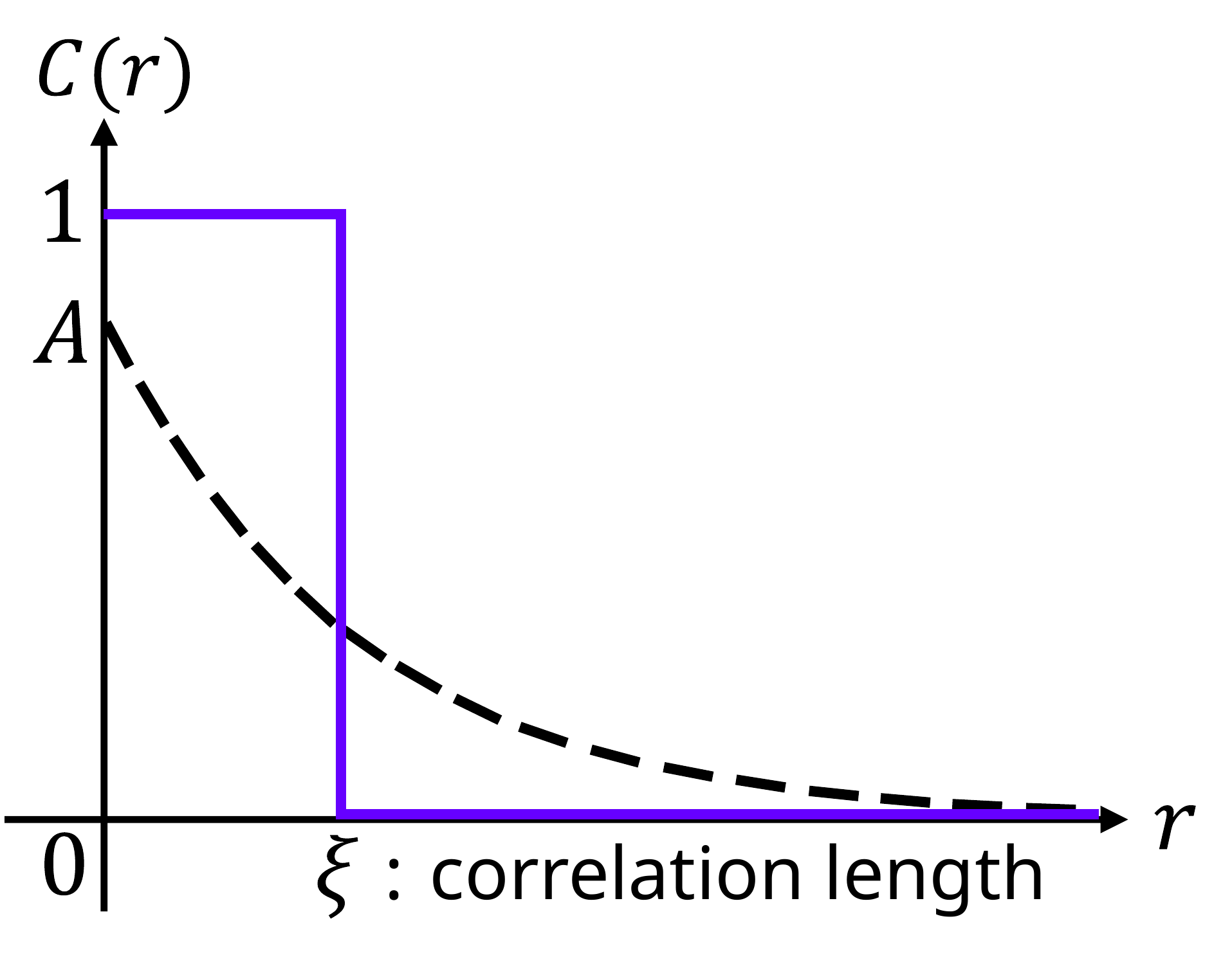}
        \caption{
        A crude approximation of the spatial correlation $C(r)$.
        The broken line is the exponential correlation,
        and the solid line the approximated step-functional correlation.
        }
    \label{fig:step_cor}
    \end{minipage}
    \hspace{10pt}
    \begin{minipage}[h]{0.45\columnwidth}
    \vspace{-9pt}
        \centering
        \includegraphics[width=5.0cm]{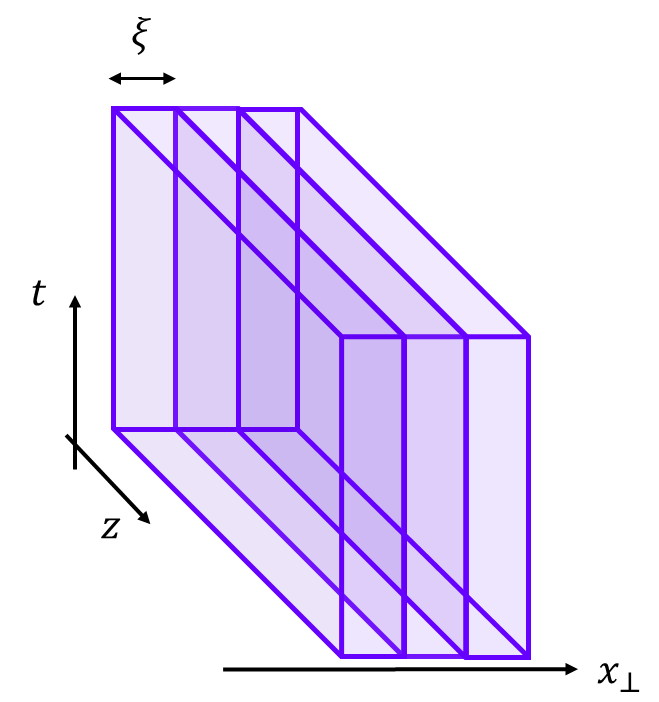}
        \caption{
        Schematic figure of DR-gauged QCD under the approximation \eqref{eq:step_cor}.
    2D QCD lives on each layer, and 2D QCDs on different layers are independent and do not interact with each other
        }
        \label{fig:tz_layer}
    \end{minipage}
\end{figure}

In this section, we consider a 2D QCD-like modeling of $tz$-projected 4D QCD in the DR gauge.
The result in the subsection \ref{subsec:correlation_temporal} shows that 
there is an exponential correlation of link-variables $U_{\mu}(s) \; (\mu = t,z)$ between two $t$-$z$ planes in the $\perp$-direction ($\perp = x, y$).
To take the effect from the correlation $C(r)$ with an analytical manner,
we make a crude approximation of replacing the correlation $C(r)$ by a step function as shown in Fig. \ref{fig:step_cor},
\begin{eqnarray}
    \label{eq:step_cor}
    C(r) 
    \to
    \theta(\xi - r)
    =
    \begin{cases}
        1 & (r < \xi) \\
        0 & (r > \xi)
    \end{cases}
\end{eqnarray}
with the correlation length $\xi$ defined in Eq. \eqref{eq:cor_length}.

This approximation means
\begin{equation}
    \label{eq:correlation_approx_in}
    C(r)= \frac{1}{N_{c}} \langle \mathrm{ReTr}~U_{t}(s) U^{\dagger}_{t}(s+ra_{\perp}) \rangle_{\mathrm{DR}}
    =
    \frac{1}{N_{c}} \langle \mathrm{ReTr}~U_{z}(s) U^{\dagger}_{z}(s+ra_{\perp}) \rangle_{\mathrm{DR}}
    =1
    \hspace{15pt}
    (r < \xi) \, ,
\end{equation}
i.e., $U_{t}(s)=U_{t}(s+ra_{\perp})$, $U_{z}(s)=U_{z}(s+ra_{\perp})$ in the shorter region of $r<\xi$. 
In contrast, one finds
\begin{equation}
    \label{eq:correlation_approx_out}
    C(r)
    =
    \frac{1}{N_{c}} \langle \mathrm{ReTr}~U_{t}(s) U^{\dagger}_{t}(s+ra_{\perp}) \rangle_{\mathrm{DR}} 
    =
    \frac{1}{N_{c}} \langle \mathrm{ReTr}~U_{z}(s) U^{\dagger}_{z}(s+ra_{\perp}) \rangle_{\mathrm{DR}}
    = 0 
    \hspace{15pt}
    (r < \xi),
\end{equation}
which physically means that $U_{t}(s)$ and $U_{t}(s+ra_{\perp})$ have no correlation in the 
$x$- and $y$-directions 
in the larger region of $r > \xi$. 

After the approximation \eqref{eq:step_cor}, $tz$-projected 4D QCD in the DR gauge can be regarded as an ensemble of 2D QCD systems on $t$-$z$ {\it layers}, which have the width of $\xi$ and are piled in the $x$- and 
$y$-directions, as shown in Fig. \ref{fig:tz_layer}.
Within each layer, $A_{t}(s)$ and $A_{z}(s)$ are uniform in the $x$- and $y$-directions, and these 2D QCD systems are independent and do not interact with each other. 

These independent $t$-$z$ layers are labeled by two integers $i$ and $j$ (instead of $x$- and $y$-coordinates), and 
the gluon fields $A_{\mu}(s)$ on the layer can be expressed as 2D-like fields,
\begin{equation}
    \label{eq:2d_A}
    \hspace{65pt}
    A_{\mu}(s)
    =
    A_{\mu}(i \xi, j \xi; t, z)
    \equiv
    A^{I}_{\mu}(t,z)
    \hspace{25pt}
    (\mu = t,z)
\end{equation}
with $I=(i, j)$.
Here, the subscript $\mu$ takes only $t$ or $z$
since $A_{\perp}(s) \; (\perp=x,y)$ are removed by the $tz$-projection \eqref{eq:tz_proj}.
Rescaling 
$A^{I}_{\mu}(t,z)$ with the correlation length $\xi$,
2D field variables are obtained,
\begin{equation}
    \label{eq:xi_A}
    \mathcal{A}^{I}_{\mu}(t,z) 
    \equiv
    \xi A^{I}_{\mu}(t,z), 
    \hspace{10pt}
    \mathcal{G}^{I}_{\mu \nu}
    \equiv
    \partial_{\mu} \mathcal{A}^{I}_{\nu}
    -
    \partial_{\mu} \mathcal{A}^{I}_{\nu}
    + 
    i \mathfrak{g} 
    \left[
        \mathcal{A}^{I}_{\mu}, \mathcal{A}^{I}_{\nu}
    \right] ,
\end{equation}
where $\mathcal{G}^{I}_{\mu \nu}$ is the 2D field strength
with the 2D gauge coupling $\mathfrak{g} \equiv g/\xi$.

Using these 2D field variables, 
the 4D action \eqref{eq:DRG_act_tzp} can be reduced to the 2D QCD action,
\begin{equation}
    \label{eq:DRG_QCD_eff_action_IR_2d_tree}
    S^{tz}_{\mathrm{DR}}
    =
    \int dt dz \:
    \sum_{I}
    \frac{1}{2}
    \mathrm{Tr}
    \left[
        \mathcal{G}^{I}_{\mu \nu}
        \mathcal{G}^{I}_{\mu \nu}
    \right]
    =
    \int dt dz \:
    \frac{1}{4}
    \delta_{I J}
    \mathcal{G}^{a, \,I}_{\mu \nu}
    \mathcal{G}^{a, \,J}_{\mu \nu} ,
\end{equation}
where the derivative interactions 
in \eqref{eq:DRG_act_tzp} have been dropped off by the approximation \eqref{eq:step_cor}.
As a general argument in 2D QCD, 
the gauge coupling has a mass dimension, and the mass scale is determined by hand.
However, in DR-gauged QCD, the mass scale of the 2D gauge coupling is automatically determined by the correlation length $\xi$ such as 
$\mathfrak{g} \equiv g/\xi$.

From this reduced 2D action  $S^{tz}_{\mathrm{DR}}$, 
the $z$-directed interquark potential on a $t$-$z$ layer is calculated at the tree-level as \cite{TS24}
\begin{equation}
    \label{eq:tree_potential2}
    V_{\mathrm{tree}}(r)
    =
    \frac{\mathfrak{g}^{2}}{2} 
    \frac{4}{3}
    r
    = 
    \sigma_{\mathrm{2D}}
    r \; ,
    \hspace{15pt}
    \sigma_{\mathrm{2D}} \equiv \frac{2}{3} \mathfrak{g}^{2}.
\end{equation}
This potential $V_{\mathrm{tree}}(r)$ is proportional to the interquark distance $r$, 
and it is a linear potential with the 2D string tension $\sigma_{\mathrm{2D}}$.
Using $g = 1.0$ (i.e., $\beta = 6/g^{2} = 6.0 $) and $\xi=0.31 \, \mathrm{fm}$,
the 2D coupling $\mathfrak{g}$ is calculated as $\mathfrak{g} \simeq 0.64 \, \mathrm{GeV}$,
and  
the 2D string tension in Eq. \eqref{eq:tree_potential2} is evaluated as
\begin{equation}
    \label{eq:2d_sigma}
    \sigma_{\mathrm{2D}}
    =
    \frac{2}{3}\mathfrak{g}^{2}
    \simeq
    1.37 \: \mathrm{GeV/fm} ,
\end{equation}
which seems to be consistent with the 4D QCD string tension $\sigma \simeq 0.89~\mathrm{GeV/fm}$
in spite of the crude approximation \eqref{eq:step_cor}.

\section{
Summary and Discussions on "Dual Graphen" and "Hologram QCD"
}
\label{sec:summary}
In this paper, 
we have studied low-dimensionalization of 4D QCD.
We have proposed a new gauge fixing of ``dimensional reduction (DR) gauge''.
Defining $tz$-projection as the replacement of $A_{x,y}(s) \to 0$ and applying it,
4D DR-gauged QCD is reduced into an ensemble of 2D QCD-like systems, which are piled in the $x$- and 
$y$-directions and interact with neighboring planes.

We have investigated low-dimensionalization properties of 4D DR-gauged QCD in $\mathrm{SU}(3)$ lattice QCD at $\beta = 6.0$.
We have found that, 
in the DR gauge, the amplitudes of two gauge components $A_{x}(s)$ and $A_{y}(s)$ are strongly suppressed, and 
these components have the large effective mass of $M_{\perp} \simeq1.7 ~ \mathrm{GeV}$, which 
indicates that $A_{x}(s)$ and $A_{y}(s)$ are inactive in the infrared region. 

Next, we have calculated the Wilson loops with the $tz$-projection, and have found that the static interquark potential is well reproduced only with $A_{t}(s)$ and $A_{z}(s)$.
In contrast, $x$- and $y$-components, $A_{t}(s)$ and $A_{z}(s)$, have been demonstrated to be inactive in the infrared region.
Then, we have proposed the possibility that low-energy phenomena in 4D DR-gauged QCD are described with 2D degrees of freedom, i.e., $A_{t}(s)$ and $A_{z}(s)$.

Focusing on the deviation between 4D $tz$-projected QCD in the DR gauge and 2D QCD,
we have calculated the spatial correlation of two temporal links,
which is induced by the deviation.
The correlation $C(r)$ has been found to decrease exponentially 
with the exponent of $m \simeq 0.64 \; \mathrm{GeV}$, which corresponds to the correlation length $\xi \equiv 1/m \simeq 0.31 \; \mathrm{fm}$. 
According to this result, we have reduced 4D DR-gauged QCD to 2D QCD with the gauge coupling of $\mathfrak{g} \equiv g/\xi$, 
and we have approximately reproduced the string tension of $\sigma_{\mathrm{2D}} \simeq 1.37 \: \mathrm{GeV/fm}$.

As a future work, it is interesting to 
consider the quark degrees of freedom. 
In the DR gauge, the gluon propagation in the $x$- and 
$y$-directions is suppressed, and gluons tend to be bounded on each $t$-$z$ plane. However, quarks are not expected to be bounded on the $t$-$z$ plane. 
In fact, 4D DR-gauged QCD would be a system where gluons (bosons) are bounded on each $t$-$z$ plane and quarks (fermions) propagate between the planes. This system is similar to the graphene, where electrons (fermions) are bounded on each 2D plane and photons (bosons) propagate between planes \cite{graphene}. Thus, the QCD system in the DR gauge can be regarded as the “dual” graphene, where roles of fermion and boson are reversed.
 
In the DR gauge, $A_t(s)$ and $A_z(s)$ are considered to be strongly correlated and propagate over long distances in the $t$- and $z$-directions, like 2D QCD on the $t$-$z$ plane. 
However, the spatial correlation $C(r)$ of two
temporal link-variables decreases exponentially,   and this means that the propagation of  $A_t(s)$ and $A_z(s)$ in the $x$- and $y$-directions is suppressed in the DR gauge. 
Thus, in the DR gauge, $A_t(s)$ and $A_z(s)$ seem to have anisotropic masses. This property seems to suggest a similarity between 4D DR-gauged QCD and 
a fracton system \cite{fracton}, where the propagation of an quasiparticle excitation is restricted in some direction. 

In the infrared region, 
we have found that 
4D QCD in the DR gauge 
is described with only 
2D gauge degrees of freedom.
This suggests that infrared essence of 4D QCD can be expressed with two-dimensional degrees of freedom. 
Then, as an interesting possibility, we propose  
an idea of “hologram QCD” 
that 4D infrared QCD is a ``holograph'' which is constructed from a ``hologram'' of the essential 2D field variables \cite{TS24}.

\section*{Acknowledgments}
We are grateful to Professor Jeff Greensite for his useful comment and discussion 
on different types of low-dimensionalization in the 2+1 Yang-Mill theory. 
H.S. is supported in part by the Grants-in-Aid for
Scientific Research [19K03869] from Japan Society for the Promotion of Science. 
K.T. is supported by Division of Graduate Studies SPRING Program at Kyoto University. 
The lattice QCD calculation has been performed by SQUID at Osaka University.

\end{document}